\documentstyle[12pt]{article}

\font\twlgot =eufm10 scaled \magstep1
\font\egtgot =eufm8
\font\sevgot =eufm7

\font\twlmsb =msbm10 scaled \magstep1
\font\egtmsb =msbm8
\font\sevmsb =msbm7

\newfam\gotfam
\def\pgot{\fam\gotfam\twlgot}
\textfont\gotfam\twlgot
\scriptfont\gotfam\egtgot
\scriptscriptfont\gotfam\sevgot
\def\got{\protect\pgot}

\newfam\msbfam
\textfont\msbfam\twlmsb
\scriptfont\msbfam\egtmsb
\scriptscriptfont\msbfam\sevmsb

\def\pBbb{\relax\ifmmode\expandafter\Bb\else\typeout{You cann't use
Bbb in text mode}\fi}
\def\Bb #1{{\fam\msbfam\relax#1}}

\def\thebibliography#1{\bigskip\section*{\large
\bf References\\}\list
   {[\arabic{enumi}]}{\settowidth\labelwidth{#1}\leftmargin\labelwidth
     \advance\leftmargin\labelsep
     \usecounter{enumi}}
     \def\newblock{\hskip .11em plus .33em minus .07em}
     \sloppy\clubpenalty4000\widowpenalty4000
     \sfcode`\.=1000\relax}

\def\op#1{\mathop{{\it\fam0} #1}\limits}

\newcommand{\beq}{\begin{equation}}
\newcommand{\eeq}{\end{equation}}
\newcommand{\ben}{\begin{eqnarray}}
\newcommand{\een}{\end{eqnarray}}
\newcommand{\be}{\begin{eqnarray*}}
\newcommand{\ee}{\end{eqnarray*}}
\newcommand{\bea}{\begin{eqalph}}
\newcommand{\eea}{\end{eqalph}}

\newcommand{\cG}{{\got g}}

\newcommand{\gj}{{\got J}}

\newcommand{\gS}{{\got S}}

\newcommand{\gF}{{\got F}}

\newcommand{\gP}{{\got P}}

\newcommand{\cP}{{\cal P}}

\newcommand{\cL}{{\cal L}}

\newcommand{\cF}{{\cal F}}

\newcommand{\cS}{{\cal S}}

\newcommand{\bL}{{\bf L}}

\newcommand{\al}{\alpha}
\newcommand{\bt}{\beta}
\newcommand{\dl}{\delta}
\newcommand{\la}{\lambda}

\newcommand{\vf}{\varphi}

\newcommand{\om}{\omega}

\newcommand{\m}{\mu}

\newcommand{\g}{\gamma}

\newcommand{\e}{\epsilon}
\newcommand{\ve}{\varepsilon}
\newcommand{\th}{\theta}

\newcommand{\si}{\sigma}

\newcommand{\w}{\wedge}

\newcommand{\ol}{\overline}

\newcommand{\dr}{\partial}

\newcommand{\ot}{\otimes}
\newcommand{\ap}{\approx}

\newenvironment{eqalph}{\stepcounter{equation}
\setcounter{equationa}{\value{equation}}
\setcounter{equation}{0}

\begin{eqnarray}}{\end{eqnarray}\setcounter{equation}{\value{equationa}}}

\newcounter{example}
\newcounter{remark}
\newcounter{theorem}
\newcounter{proposition}
\newcounter{lemma}
\newcounter{corollary}
\newcounter{definition}

\def\thedefinition{\arabic{definition}}

\newcommand{\mar}[1]{}

\begin{document}
\hbox{}

\begin{center}

{\large \bf Noether conservation laws in
higher-dimensional  Chern--Simons theory}
\bigskip

{\sc Giovanni Giachetta},\footnote{E-mail:
giovanni.giachetta@unicam.it} {\sc Luigi
Mangiarotti}\footnote{E-mail: luigi.mangiarotti@unicam.it}

{\small \it Department of Mathematics and Informatics,
University of Camerino, 62032 Camerino (MC) Italy}
\medskip

{\sc Gennadi Sardanashvily}\footnote{E-mail:
sard@grav.phys.msu.su}

{\small \it
Department of Theoretical Physics, Moscow State University, 117234
Moscow, Russia}

\end{center}

\begin{small}

\noindent
Though a global Chern--Simons $(2k-1)$-form is not
gauge-invariant, this form seen as a
Lagrangian of higher-dimensional gauge theory
leads to the conservation law
of a modified Noether current.
\end{small}

\bigskip\bigskip

One usually considers Chern--Simons (henceforth CS) gauge
theory on
a principal bundle over a three-dimensional manifold
whose  Lagrangian is the local
CS form derived from the local
transgression formula for the second Chern characteristic form.
This Lagrangian fails to be globally defined, unless a
principal bundle is trivial (e.g., if its structure group is
simply connected \cite{fre}). Though the local CS Lagrangian is
not gauge-invariant, it leads to the (local) conservation
law of the  modified
Noether current \cite{bor98,book,sard97}. This
result is extended  to the global three-dimensional CS theory
\cite{al,bor2}.
Its Lagrangian is well defined, but depends on a
background gauge
potential. Therefore, it is gauge-covariant, but not
gauge-invariant. At the same time, the corresponding
Euler--Lagrange operator  is gauge-invariant, and the above
mentioned  gauge conservation law takes place. We aim to show
that any higher-dimensional CS theory admits such a
conservation law.

There are different approaches to the study of Lagrangian
conservation laws. We use the so called first variational
formula, which enables one to obtain
conservation laws if a symmetry is broken
\cite{book,book00,sard97}.

Let us consider a first order
field theory on a fibre bundle $Y\to X$ over an $n$-dimensional
smooth manifold $X$. Its configuration space is
the first order
jet manifold $J^1Y$ of sections of $Y\to X$.
Given bundle coordinates $(x^\la,y^i)$ on a fibre bundle $Y\to
X$, its first and second order jet manifolds $J^1Y$ and
$J^2Y$ are endowed with the
adapted coordinates $(x^\la,y^i,y^i_\m)$ and $(x^\la,y^i,y^i_\m,
y^i_{\la\m})$, respectively.
One can think of $y^i_\m$ and $y^i_{\la\m}$ as being coordinates
of first and second derivatives of dynamic variables. We
use the notation
$\om=d^nx$ and
$\om_\la=\dr_\la\rfloor \om$.

A first order Lagrangian of field theory on $Y\to X$ is defined
as a density
\mar{a1}\beq
L=\cL(x^\m,y^j,y^j_\m) \om \label{a1}
\eeq
on the first order jet manifold
$J^1Y$ of $Y\to X$. Given a Lagrangian $L$ (\ref{a1}), the
corresponding Euler--Lagrange operator reads
\mar{305}\beq
\dl L= \dl_i\cL \th^i\w\om=(\dr_i\cL-
d_\la\dr^\la_i)\cL \th^i\w\om,
\label{305}
\eeq
where $\th^i=dy^i- y^i_\la dx^\la$ are contact forms
and
\be
d_\la=\dr_\la
+y^i_\la\dr_i +y^i_{\la\m}\dr_i^\m
\ee
are the total
derivatives, which yield the total differential
$d_H\vf=dx^\la\w d_\la\vf$ acting on exterior forms on $J^1Y$.
The kernel
Ker$\,\dl L\subset J^2Y$ of the Euler--Lagrange operator (\ref{305})
defines the Euler--Lagrange
equations
\mar{a2}\beq
\dl_i\cL=(\dr_i\cL- d_\la\dr^\la_i)\cL=0. \label{a2}
\eeq
A Lagrangian $L$ (\ref{a1}) is said to be variationally
trivial if
$\dl L=0$. This property holds iff
$L=h_0(\vf)$, where $\vf$ is a closed $n$-form on $Y$ and
$h_0$ is the horizontal projection
\be
h_0(dx^\la)=dx^\la, \qquad h_0(dy^i)=y^i_\la dx^\la, \qquad h_0(dy^i_\m)=
y_{\la\m}^idx^\la.
\ee
The relation $d_H\circ h_0=h_0\circ d$ holds.

To obtain Noether conservation laws, one considers
local one-parameter groups of vertical bundle automorphisms
(gauge transformations) of
$Y\to X$. Their infinitesimal
generators are vertical vector fields
$u=u^i(x^\m,y^j)\dr_i$
on $Y\to X$ whose prolongation onto
$J^1Y$ reads
\mar{1.21}\beq
J^1u=u^i\dr_i +d_\la u^i\dr^\la_i.
\label{1.21}
\eeq
A Lagrangian $L$ is invariant under a one-parameter group of
gauge transformations generated by a vector field $u$ iff its
Lie derivative
\mar{r8}\beq
\bL_{J^1u}L=J^1u\rfloor dL=(u^i\dr_i\cL +d_\la
u^i\dr^\la_i\cL)\om \label{r8}
\eeq
along
$J^1u$ vanishes.
The first variational
formula provides the canonical decomposition
\mar{bC30'}\beq
\bL_{J^1u}L=
  u\rfloor \dl L + d_H(u\rfloor H_L) = u^i\dl_i\cL\om +
d_\la(u^i\dr^\la_i\cL)\om,
\label{bC30'}
\eeq
where $H_L=\cL\om +\dr_i^\la\cL\th^i\w\om_\la$
is the Poincar\'e--Cartan form of $L$, and
\mar{Q30}\beq
\gj_u =u\rfloor H_L=\gj_u^\la\om_\la
=u^i\dr^\la_i\cL\om_\la
\label{Q30}
\eeq
is the symmetry current along $u$. On the shell
(\ref{a2}),  the first variational formula (\ref{bC30'}) leads
to the weak  equality
\mar{J4}\beq
\bL_{J^1u}L\ap-d_H\gj_u, \qquad
u^i\dr_i\cL +d_\la u^i\dr^\la_i\cL \ap
d_\la(u^i\dr^\la_i\cL). \label{J4}
\eeq
If $\bL_{J^1u}L=0$, we obtain
the Noether conservation law
\mar{a4}\beq
0\ap d_H\gj_u \label{a4}
\eeq
of the symmetry current $\gj_u$ (\ref{Q30}).
If the Lie derivative (\ref{r8}) reduces to
the total differential
\mar{r4}\beq
\bL_{J^ru}L=d_H\si, \label{r4}
\eeq
then the weak equality (\ref{J4}) takes the form
\mar{r5}\beq
0\ap d_H(\gj_u-\si), \label{r5}
\eeq
regarded as a conservation law of the modified symmetry
current
$\ol\gj=\gj_u-\si$.

Now, let us turn to gauge theory of principal connections on a
principal bundle $P\to X$ with a structure
Lie group $G$. Let $J^1P$ be the first order jet
manifold of $P\to X$
and
\mar{r43}\beq
C=J^1P/G\to X \label{r43}
\eeq
the quotient of $P$ with respect to the canonical action of
$G$ on
$P$ \cite{book00,sard97}.
There is one-to-one correspondence between the principal
connections on
$P\to X$ and the sections of the fibre bundle $C$
(\ref{r43}), called the
connection bundle. Given an atlas $\Psi$ of $P$, the
connection bundle
$C$ is provided with bundle
coordinates $(x^\la, a^r_\m)$ such that, for any its section
$A$, the
local functions $A^r_\m=a^r_\m\circ A$ are coefficients of the
familiar local connection form. From the physical viewpoint,
$A$
is a gauge potential.

  The infinitesimal
generators of one-parameter groups of gauge transformations of
the principal bundle $P$ are $G$-invariant vertical vector
fields on $P$. There is one-to-one correspondence between these
vector fields and the sections of the quotient
$V_GP=VP/G\to X$
of the vertical tangent bundle $VP$ of $P\to X$ with respect to the
canonical action of $G$ on $P$. The typical fibre of
$V_GP$ is the right Lie algebra $\cG$ of the Lie group $G$,
acting
on this typical fibre by the adjoint representation. Given an
atlas
$\Psi$ of $P$ and a basis $\{\e_r\}$ for the Lie algebra
$\cG$, we
obtain the
fibre bases $\{e_r\}$ for $V_GP$. If $\xi=\xi^pe_p$ and
$\eta=\eta^q e_q$ are sections of $V_GP\to X$, their bracket is
\be
[\xi,\eta]=c^r_{pq}\xi^p\eta^q e_r,
\ee
where $c^r_{pq}$ are the structure constants of $\cG$.
Note that the connection bundle $C$ (\ref{r43}) is an affine bundle
modelled over the vector bundle $T^*X\ot V_GP$, and elements of $C$ are
represented by local $V_GP$-valued 1-forms $a^r_\m
dx^\m\ot e_r$.
The infinitesimal generators of gauge transformations of
the connection bundle $C\to X$ are vertical vector fields
\mar{r47}\beq
\xi_C=(\dr_\m\xi^r +c^r_{pq}a^p_\m\xi^q)\dr^\m_r. \label{r47}
\eeq

The connection bundle $C\to X$ admits the canonical $V_GP$-valued 2-form
\mar{r60}\beq
\gF=(da^r_\m\w dx^\m +\frac12 c^r_{pq}a^p_\la a^q_\m dx^\la\w dx^\m)\ot
e_r, \label{r60}
\eeq
which is the curvature of the canonical connection on the principal
bundle $C\times P\to C$ \cite{book00}.
Given a section $A$ of $C\to X$, the pull-back
\mar{r47'}\beq
F_A=A^*\gF=\frac12 F^r_{\la\m}dx^\la\w dx^\m\ot e_r,
\qquad
F^r_{\la\m}=\dr_\la A^r_\m-\dr_\m A^r_\la +c^r_{pq}A^p_\la
A^q_\m,
\label{r47'}
\eeq
of $\gF$ onto $X$ is the strength form of a gauge potential $A$.

Turn now to the CS forms.
Let $I_k(\e)=b_{r_1\ldots r_k}\e^{r_1}\cdots \e^{r_k}$ be a $G$-invariant
polynomial of degree $k>1$ on the Lie algebra $\cG$ written with
respect to its basis $\{\e_r\}$, i.e.,
\be
\op\sum_j b_{r_1\ldots r_k}\e^{r_1}\cdots
c^{r_j}_{pq}\e^p\cdots
\e^{r_k}=
kb_{r_1\ldots r_k}c^{r_1}_{pq}\e^p\e^{r_2}\cdots \e^{r_k}=0.
\ee
Let us associate to $I(\e)$ the closed gauge-invariant
$2k$-form
\mar{r61}\beq
P_{2k}(\gF)=b_{r_1\ldots r_k}\gF^{r_1}\w\cdots\w \gF^{r_k} \label{r61}
\eeq
on $C$. Let $A$ be a section of $C\to X$. Then, the pull-back
\mar{r63}\beq
P_{2k}(F_A)=A^*P_{2k}(\gF) \label{r63}
\eeq
of $P_{2k}(\gF)$ is a closed characteristic form on $X$. Recall that
the de Rham cohomology of $C$ equals that of $X$ since $C\to X$
is an affine bundle. It follows that $P_{2k}(\gF)$ and $P_{2k}(F_A)$ possess
the same
cohomology class
\mar{r62}\beq
[P_{2k}(\gF)]=[P_{2k}(F_A)] \label{r62}
\eeq
for any principal connection $A$. Thus, $I_k(\e)\mapsto
[P_{2k}(F_A)]\in H^*(X)$
is the familiar Weil homomorphism.

Let $B$ be a fixed section of the connection bundle $C\to X$.
Given the characteristic form $P_{2k}(F_B)$ (\ref{r63}) on $X$, let the same
symbol stand for its pull-back onto $C$. By virtue of the equality
(\ref{r62}), the difference $P_{2k}(\gF)-P_{2k}(F_B)$ is an exact form on $C$.
Moreover, similarly to the well-known transgression formula on a
principal bundle $P$, one can obtain the following transgression formula
on $C$:
\mar{r64,5}\ben
&& P_{2k}(\gF)-P_{2k}(F_B)=d\gS_{2k-1}(B), \label{r64}\\
&&  \gS_{2k-1}(B)=k\op\int^1_0 \gP_{2k}(t,B)dt, \label{r65}\\
&& \gP_{2k}(t,B)=b_{r_1\ldots r_k}(a^{r_1}_{\m_1}-B^{r_1}_{\m_1})dx^{\m_1}\w
\gF^{r_2}(t,B)\w\cdots \w \gF^{r_k}(t,B),\nonumber\\
&& \gF^{r_j}(t,B)=
[d(ta^{r_j}_{\m_j} +(1-t)B^{r_j}_{\m_j})\w dx^{\m_j} +\nonumber\\
&& \qquad \frac12c^{r_j}_{pq}
(ta^p_{\la_j} +(1-t)B^p_{\la_j})(ta^q_{\m_j} +(1-t)B^q_{\m_j})dx^{\la_j}\w
dx^{\m_j}]\ot e_r.
\nonumber
\een
Its pull-back by means of a section $A$ of $C\to X$ gives
the transgression formula
\be
P_{2k}(F_A)-P_{2k}(F_B)=d S_{2k-1}(A,B)
\ee
on $X$. For instance, if $P_{2k}(F_A)$ is the characteristic Chern $2k$-form,
then $S_{2k-1}(A,B)$ is the familiar CS $(2k-1)$-form.
Therefore, we agree to call $\gS_{2k-1}(B)$
(\ref{r65}) the CS form on the connection bundle $C$.
In particular, one can choose the local section $B=0$.
Then, $\gS_{2k-1}=\gS_{2k-1}(0)$ is
the local CS form. Let $S_{2k-1}(A)$ denote its pull-back
onto $X$ by means of a section $A$ of $C\to X$. Then, the CS form
$\gS_{2k-1}(B)$ admits the decomposition
\mar{r75}\beq
\gS_{2k-1}(B)=\gS_{2k-1} -S_{2k-1}(B) +dK_{2k-1}(B). \label{r75}
\eeq

Let $J^1C$ be the first order jet manifold of the
connection bundle $C\to X$ equipped with the adapted coordinates
$(x^\la, a^r_\m, a^r_{\la\m})$.
Let us consider the pull-back of the CS form (\ref{r65}) onto $J^1C$
denoted by the same symbol $\gS_{2k-1}(B)$, and let
\mar{r74}\beq
\cS_{2k-1}(B)=h_0 \gS_{2k-1}(B) \label{r74}
\eeq
be its horizontal projection.
This is given by the formula
\be
&& \cS_{2k-1}(B)=k\op\int^1_0 \cP_{2k}(t,B)dt, \\
&& \cP_{2k}(t,B)=b_{r_1\ldots r_k}(a^{r_1}_{\m_1}-B^{r_1}_{\m_1})dx^{\m_1}\w
\cF^{r_2}(t,B)\w\cdots \w \cF^{r_k}(t,B),\\
&& \cF^{r_j}(t,B)= \frac12[
ta^{r_j}_{\la_j\m_j} +(1-t)\dr_{\la_j}B^{r_j}_{\m_j}
- ta^{r_j}_{\m_j\la_j} -(1-t)\dr_{\m_j}B^{r_j}_{\la_j})+\\
&& \qquad \frac12c^{r_j}_{pq}
(ta^p_{\la_j} +(1-t)B^p_{\la_j})(ta^q_{\m_j} +(1-t)B^q_{\m_j}]dx^{\la_j}\w
dx^{\m_j}\ot e_r.
\ee

Now, let us consider the CS gauge model on a $(2k-1)$-dimensional base
manifold $X$ whose Lagrangian
\mar{r80}\beq
L_{\rm CS}=\cS_{2k-1}(B) \label{r80}
\eeq
is the CS form (\ref{r74}) on $J^1C$. Clearly, this Lagrangian is not
gauge-invariant.
Let $\xi_C$ (\ref{r47}) be the infinitesimal generator of gauge
transformations of the connection bundle $C$.
Its jet prolongation onto $J^1C$ is
\be
J^1\xi_C=\xi^r_\m\dr^\m_r + d_\la\xi^r_\m\dr^{\la\m}_r.
\ee
  The Lie derivative of the Lagrangian
$L_{\rm CS}$ along $J^1\xi_C$ reads
\mar{r81}\beq
\bL_{J^1\xi_C} \cS_{2k-1}(B)= J^1\xi_C\rfloor d(h_0\gS_{2k-1}(B))
=\bL_{J^1\xi_C}(h_0 \gS_{2k-1}(B)).
\label{r81}
\eeq
A direct computation shows that
\be
&&\bL_{J^1\xi_C}(h_0
\gS_{2k-1}(B))=h_0(\bL_{\xi_C}\gS_{2k-1}(B)) =h_0(\xi_C\rfloor
d\gS_{2k-1}(B)+ d(\xi_C\rfloor\gS_{2k-1}(B))=\\
&& \qquad h_0(\xi_C\rfloor
d\gS_{2k-1}(B) +d_H(h_0(\xi_C\rfloor\gS_{2k-1}(B)).
\ee
By virtue of the transgression formula (\ref{r64}), we have
\be
d(\xi_C\rfloor d\gS_{2k-1}(B))=\bL_{\xi_C}(d\gS_{2k-1}(B))=
\bL_{\xi_C}P_{2k}(\gF)=0.
\ee
It follows that $\xi_C\rfloor d\gS_{2k-1}(B)$ is a closed form on $C$,
i.e.,
\be
\xi_C\rfloor d\gS_{2k-1}(B)=d\psi + \vf,
\ee
where $\vf$ is a non-exact $(2k-1)$-form on $X$. Moreover, $\vf=0$ since
$P(\gF)$, $k>1$, does not contain terms linear in $da^r_\m$.
Hence, the Lie derivative (\ref{r81}) takes the form (\ref{r4}) where
\be
\bL_{J^1\xi_C} \cS_{2k-1}(B)=d_H \si,
\qquad \si=h_0(\psi + \xi_C\rfloor\gS_{2k-1}(B)).
\ee
As a consequence,  CS theory with the Lagrangian (\ref{r80})
admits the conservation law (\ref{r5}).

In a more general setting, one can
consider the sum of the CS Lagrangian (\ref{r80}) and some
gauge-invariant Lagrangian.
For instance, let $G$ be a semi-simple group and $a^G$
the Killing form on $\cG$. Let
\mar{r48}\beq
P(\gF)=\frac{h}{2}a^G_{mn}\gF^m\w \gF^n \label{r48}
\eeq
be the second Chern form up to a constant multiple. Given a section $B$ of
$C\to X$, the transgression formula
(\ref{r64}) on $C$ reads
\mar{r49}\beq
P(\gF)-P(F_B)=d\gS_3(B), \label{r49}
\eeq
where $\gS_3(B)$ is the CS 3-form up to a constant multiple.
Let us consider the gauge model on a 3-dimensional base
manifold whose Lagrangian is the sum
$L= L_{\rm CS}+L_{\rm inv}$
of the CS Lagrangian
\mar{r50}\ben
&&L_{\rm CS}=h_0(\gS_3(B))=
[\frac12ha^G_{mn} \ve^{\al\bt\g}a^m_\al(\cF^n_{\bt\g} -\frac13
c^n_{pq}a^p_\bt a^q_\g)  \label{r50}\\
&& \qquad -\frac12ha^G_{mn}
\ve^{\al\bt\g}B^m_\al(F(B)^n_{\bt\g} -\frac13 c^n_{pq}B^p_\bt
B^q_\g) -d_\al(ha^G_{mn} \ve^{\al\bt\g}a^m_\bt B^n_\g)]d^3x,
\nonumber\\
&& \cF=h_0\gF=\frac12 \cF^r_{\la\m}dx^\la\w
dx^\m\ot e_r, \qquad
\cF^r_{\la\m}=a^r_{\la\m}-a^r_{\m\la} +c^r_{pq}a^p_\la a^q_\m,
\nonumber
\een
and some gauge-invariant Lagrangian
\mar{r85}\beq
L_{\rm inv}= \cL_{\rm
inv}(x^\la,a^r_\m,a^r_{\la\m},z^A,z^A_\la)d^3x \label{r85}
\eeq
of gauge potentials $a$ and matter fields $z$. Then, the first
variational formula (\ref{bC30'}) on-shell takes the form
\mar{r86}\beq
\bL_{J^1\xi_C}L_{\rm CS}\ap d_H(\gj_{\rm CS} +\gj_{\rm inv}), \label{r86}
\eeq
where $\gj_{\rm CS}$ is the Noether current of the CS Lagrangian
(\ref{r50}) and $\gj_{\rm inv}$ is that of the gauge-invariant
Lagrangian (\ref{r85}). A simple calculation gives
\be
&& \bL_{J^1\xi_C}L_{\rm
CS}=-d_\al(ha^G_{mn}\ve^{\al\bt\g}(\dr_\bt\xi^m a^n_\g +
(\dr_\bt\xi^m +c^m_{pq}a^p_\bt\xi^q) B^n_\g))d^3x,\\
&& \gj_{\rm CS}^\al=ha^G_{mn}\ve^{\al\bt\g}(\dr_\bt\xi^m
+c^m_{pq}a^p_\bt\xi^q)(a^n_\g- B^n_\g).
\ee
Substituting these expressions into the weak equality
(\ref{r86}), we come to the conservation law
\be
0\ap d_\al[ha^G_{mn}\ve^{\al\bt\g}(2\dr_\bt\xi^m a^n_\g +
c^m_{pq}a^p_b a^n_\g\xi^q)+ \gj_{\rm inv}^\al]
\ee
of the modified Noether current
\be
\ol\gj=ha^G_{mn}\ve^{\al\bt\g}(2\dr_\bt\xi^m a^n_\g +
c^m_{pq}a^p_b a^n_\g\xi^q)+ \gj_{\rm inv}^\al.
\ee

\end{document}